\documentclass[11pt]{article}
\usepackage{geometry}                
\usepackage[parfill]{parskip}    
\usepackage{subcaption}
\usepackage{graphicx}
\usepackage{caption}

\usepackage{amsmath, amsthm, amsfonts}
\usepackage{amssymb}

\usepackage{tikz-cd}
\usetikzlibrary{positioning}
\usepackage{caption}
 
\newcommand{\sect}[1]{\setcounter{equation}{0}\section{#1}}

\usepackage{color}
     \usepackage[colorlinks]{hyperref}
\hypersetup{linkcolor=blue,%
citecolor=red,%
urlcolor=cyan}
\usepackage{url} 

\begin{document}

\title{Expected values for SUSY hierarchies of
\\ 
Jaynes-Cummings Hamiltonians}

\author{
\.{I}smail Burak Ate\c s$^a$\footnote{ismailburakates@gmail.com, ORCID: \href{http://orcid.org/0000-0001-8262-2572}{0000-0001-8262-2572}}, 
\c Seng\"ul Kuru$^a$\footnote{kuru@science.ankara.edu.tr-Corresponding author, ORCID: \href{http://orcid.org/0000-0001-6380-280X}{0000-0001-6380-280X}}, 
Javier Negro$^b$\footnote{jnegro@uva.es, 
ORCID: \href{http://orcid.org/0000-0002-0847-6420}{0000-0002-0847-6420}},
Ege \"Ozkan$^c$\footnote{
e.ozkan@bilkent.edu.tr,
ORCID: \href{http://orcid.org/0009-0005-0944-4665}{0009-0005-0944-4665}
}
\\ 
\\
$^a$Department of Physics, Faculty of Science, Ankara University,\\ 06100 Ankara, T\"urkiye\\
$^b$Departamento de F\'i{}sica Te\'orica, At\'omica y \'Optica and IMUVA,\\Universidad de Valladolid, 47011 Valladolid, Spain
\\
$^c$Bilkent University, Department of Physics, 06800 Ankara, T\"urkiye
}

\maketitle

\begin{abstract}

The aim of this letter is to compute the evolution of some expected values, such as the field operators $a^\pm$, quadratures and atomic inversion,
under supersymmetric (SUSY) partner Hamiltonians associated to the Jaynes-Cummings Hamiltonian of quantum optics. This kind of SUSY partners are characterized by having spectra which differ in a finite number of energy levels. We wish to elucidate if the partner connection has any influence on these expected values. In particular, we want also to know in which way the classical and revival times are affected by such SUSY partners.

\end{abstract}

\sect{Introduction}

In a recent work \cite{ates25}, we have shown that a sequence of  Jaynes-Cummings (JC) Hamiltonians (quantum Rabi models in therotating wave approximation (RWA)) \cite{jaynes63,Haroche96,gerry04,Larson24} with certain detuning parameters but the same coupling coefficient, have the property of being supersymmetric (SUSY) partners. 
In other words, these Hamiltonians are intertwined by operators and they share the same spectrum except for a finite number of eigenvalues. The quantum Rabi model has different coupling regimes  \cite{Rossatto17,Solano19,Huang20}, but in principle SUSY transformations can be implemented  under the restriction of the RWA approximation.

SUSY transformations are well known in quantum mechanics connecting pairs of Schr\"odinger Hamiltonians \cite{fernandez10,cooper95,matveev91,junker96}. Two Hamiltonians are SUSY partners, in a wide sense, if they are intertwined by a  differential operator. Then, they will have the same spectrum up to eigenvalues of eigenfunctions annihilated by the intertwining operator  \cite{cooper95,junker96}. This property may appear in different contexts and forms such as the factorization method \cite{infeld51}, Darboux transformations \cite{matveev91}, or supersymmetric Hamiltonians  \cite{fernandez10,cooper95,matveev91,junker96}. If  this type of SUSY transformations can be applied in iterative way to a sequence of Hamiltonians $H_n$, $n\in \mathbb Z$, then we may arrive to a sequence of shape invariant potentials $V_n$. That is, potentials having the same formal expression which depend on the parameter $n$. Sometimes, this is referred as a Hamiltonian hierarchy \cite{cooper95}. 

More recently, SUSY methods have been applied to Dirac Hamiltonians of two dimensional materials such as graphene \cite{kuru21}. Their extension to the interaction matter-radiation has been considered previously as a part of supersymmetry \cite{negro04,hussin06,negro24,lara24,alhaidari06,panahi15,lara13,lara21,mubark23,alexio07,castanos13}. However, this is really a new area where up to now 
these methods were not fully  developed in the same style as in the above well known situations. In a previous work \cite{ates25}, we have shown that this is possible.
We computed a sequence of JC Hamiltonians intertwined by matrix operators to form a shape invariant sequence, or hierarchy, similar to those characterized, for instance, in quantum mechanics. In fact, the application of SUSY techniques in JC Hamiltonians is similar to that in graphene. These results confirm what it is already widely accepted: that there are many links between the Hamiltonians used in condensed matter and  quantum optics which makes possible to apply  similar methods in both fields.

JC Hamiltonians have interesting properties such as collapses and revivals \cite{rempe87,Chong25} showing the quantum nature of this model.  The purpose of this letter is to compute some expected values of  operators in the evolution of states under these SUSY partner Hamiltonians to show if the collapses and revivals have an special resemblance when they belong to the same SUSY hierarchy or not. In other words, if the fact that two JC Hamiltonians are SUSY partners, have a relevance on their measurable expected values.  Tensor products of coherent states (which are more appropriate than Fock states) and atomic or qubit states, will be used to build the initial evolution states  \cite{gerry04}. Since SUSY methods lead us to similar spectrum for all Hamiltonians in the hierarchy, it will be easy to obtain the characteristic classical period and revival times of the evolution states \cite{robinett04}.

This article is a continuation of the article before that was related with the SUSY hierarchies of JC Hamiltonian systems \cite{ates25}. Here, we will consider  expected values of $\sigma_3$ regarding the atomic inversion \cite{gerry04,Larson24}, and those of pure field operators such as the creation and annihilation $a^\pm$ of monochromatic  radiation field (or their quadrature operators), to study if some type of similarity can be interpreted as  the fingerprints for being in the same SUSY hierarchy. However, as we will see, sometimes this feature may be not so specific of SUSY partner Hamiltonians and it may have a broader origin. 

The organization of this letter is as follows. In section 2, we summarise the notation and results on the SUSY hierarchies of JC Hamiltonians. 
In particular, we stress on the particular  parameter values characteristics of partner Hamiltonians and how are their spectrum and eigenfunctions.
The following sections will be devoted to compute the evolution of expected values for an atomic inversion operator $\sigma_3$ (section 3), for  $a^\pm$ and quadrature field operators (section 4) under SUSY partner Hamiltonians.
The main results and further remarks are supplied in the final section. 

\sect{SUSY partners of JC Hamiltonian}
We will work with SUSY partners of the JC Hamiltonian that
were constructed in a previous paper  \cite{ates25}. Here, we will introduce them in a direct way so that their main properties be immediately explicit. Firstly, we recall the solutions of the standard JC Hamiltonian.

\subsection{Physical solutions of the JC Hamiltonian}

 We will consider the JC Hamiltonian  $H_{\rm JC}$ corresponding to the Rabi quantum model in the RWA  which describes the interaction of a two level atom with a single mode radiation field \cite{ates25}. 
The Hilbert space of the JC Hamiltonian is generated by the tensor product of the two-dimensional atomic space generated by the ground and excited states 
$|g\rangle$, $|e\rangle$, respectively, and the radiation quantum space of frequency $\omega$, generated by the number states $|n\rangle:=\psi_n$.
We make use of the standard notation for tensor products involving these ground and excited states:
\begin{equation}
|g\rangle \otimes \psi_n :=  \left(\begin{array}{c}
0
\\
{\psi_n}
\end{array}\right)
,\,\quad
|e\rangle \otimes {\psi_n}:= \left(\begin{array}{c}
{\psi_n}
\\
0
\end{array}\right) 
\end{equation}

The total space can be expressed as the direct sum of these two subspaces. Then, the JC Hamiltonian, in appropriate units, takes the following  expression
\begin{equation}\label{H3m}
H_{\rm JC}:=
\left(
\begin{array}{cc}  
a^- a^+ +\delta &\lambda a^-
\\[1ex]
\lambda a^+ & a^+a^- -\delta
\end{array}\right) 
\end{equation}
where the {dimensionless} detuning parameter $\delta$ is
\begin{equation}\label{param}
\delta = 
\frac{\omega_0-\omega}{2\omega}
\end{equation} 
\begin{itemize}
\item
The energy between the two atomic levels is denoted by $\hbar \omega_0$ while $\hbar \omega$ will be that  of the radiation field.
\item
{The coefficient $\lambda$ is proportional to the coupling of radiation-atom and it is dimensionless.}

\item
The operators $a^\pm$ are creation/annihilation of a radiation quantum
$\hbar \omega$. They satisfy standard commutation relation:
\begin{equation}
[a^-,a^+] = 1
\end{equation}
In terms of these operators the number operator has the form $N=a^+a^-$ and it satisfies
$N \psi_{n}= n \psi_{n},\quad n= 0,1,2,\dots$ The number state $|n\rangle$ or $\psi_n$ represents a $n$-photon state with energy $n \hbar\omega$. The space of quantized radiation in this way is the Fock space having the number states basis.
\end{itemize} 

\begin{figure}[h!]
\begin{center}
\includegraphics[width=0.45\textwidth]{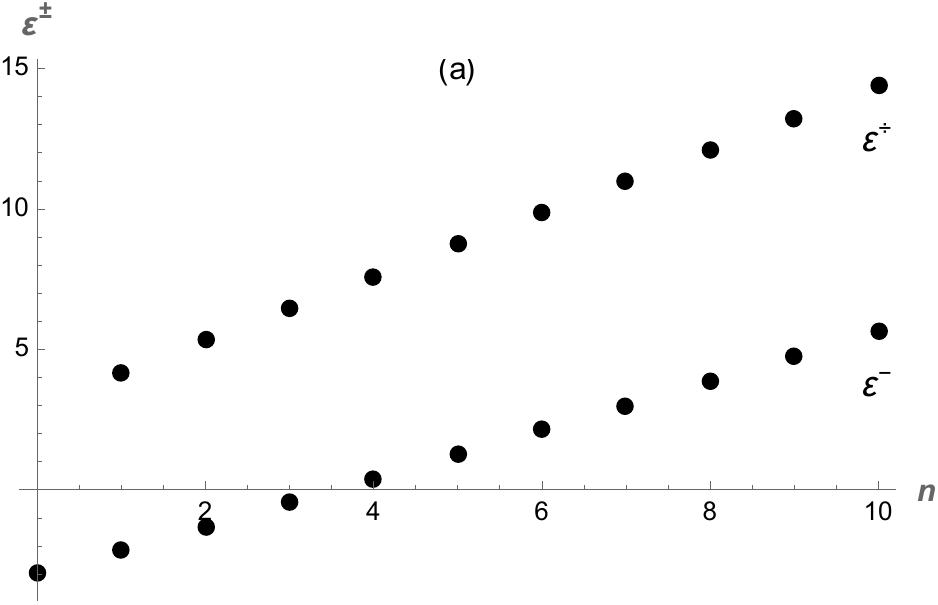}
\qquad\quad
\includegraphics[width=0.45\textwidth]{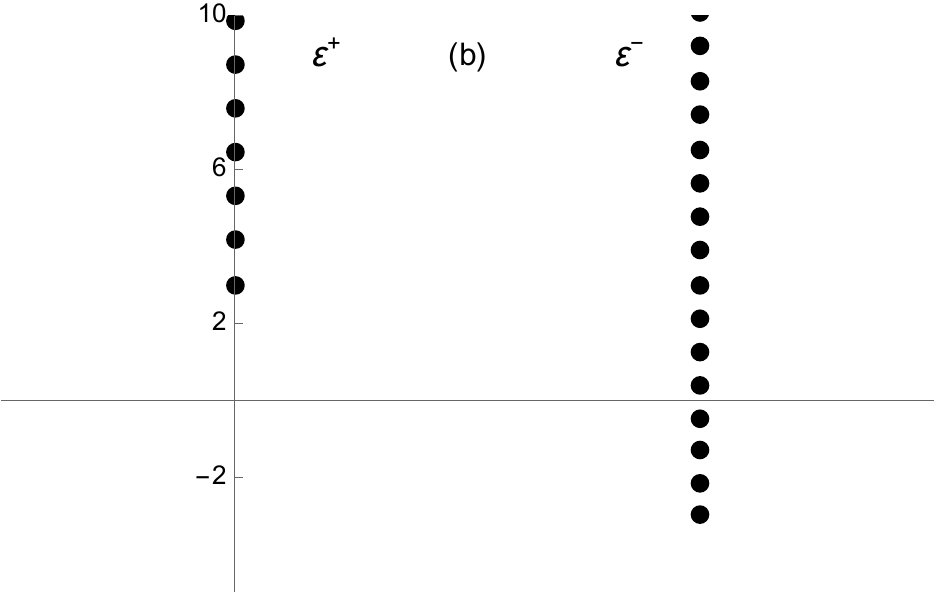}
\end{center}
\caption{\small Two distinct ways to represent the lower energy levels for the JC Hamiltonian (with
$\delta=3$, $\lambda=1$). (a) Represents the pairs of eigenvalues $\varepsilon_n^\pm$ depending on $n$, 
 and (b) represents two columns, the left corresponding to $\varepsilon_n^+$,
 the right to $\varepsilon_n^-$ eigenvalues, ordered by their values.
\label{fig1}}
\end{figure}

The JC eigenvalue equation is
\begin{equation}
H_{\rm JC}\Psi = \varepsilon \Psi,\qquad \varepsilon= \frac{E}{\hbar \omega}
\end{equation}
{The JC Hamiltonian has a symmetry, known as excitation number
\begin{equation}\label{ven}
N_{\rm e} =\left(\begin{array}{ll}
a^- a^+& 0
\\[1ex]
0 & a^+a^-
\end{array}\right)
\end{equation}
This symmetry has eigenvalues $n = 0,1,\dots$, and the corresponding eigenspaces are $V_{n}$ with dimensions ${\rm dim} V_0=1$, ${\rm dim} V_n= 2$, for $n\neq 0$.}
Then, the physical eigenstates  (they are  two-component spinors) of the JC Hamiltonian  ({\ref{H3m}}) belong to the
 these subspaces \cite{gerry04}
\begin{equation}\label{vn}
V_n =\left(\begin{array}{l}
c_1 \psi_{n-1}
\\[1ex]
c_2\psi_n
\end{array}\right)\,,
\qquad c_1,c_2\in \mathbb C,\quad n=0,1,2,\dots
\end{equation}
Thus, the eigenvalues and (not normalized) eigenfunctions are as follows
\cite{gerry04} 
\begin{equation}\label{phys}
\begin{array}{lll}
\varepsilon_n^\pm = n \pm \sqrt{\delta^2 +n \lambda^2},\quad 
&\Psi_n^\pm = \left(\begin{array}{c}
(\delta\pm \sqrt{\delta^2 +n \lambda^2})\,\psi_{n-1}
\\[1ex]
\sqrt{n}\,\lambda\, \psi_n
\end{array}\right),\quad &n=1,2,\dots
\\[4.5ex]
\varepsilon^-_0 = -\delta,\qquad
&\Psi^-_0 = \left(\begin{array}{c}
0
\\[1ex]
 \psi_0
\end{array}\right),\quad &n=0
\end{array} 
\end{equation}
For the positive values $n=1,2,\dots$ there are two types of excited solutions $\Psi_n^\pm$, while only one ground single state, $\Psi^-_0$, is obtained for $n=0$, as can be seen in Fig.~\ref{fig1}. These solutions are called dressed states. 

\subsection{SUSY partners}

In order to introduce the SUSY partner Hamiltonians, let us pay attention to the energy eigenvalue formula for
$\varepsilon_n^\pm$ of (\ref{phys}). We notice that if in that formula we apply the following modifications: 

\noindent
(i)  change the detuning parameter
$\delta^2$ by $\delta^2+\lambda^2$ 

\noindent
(ii)  add one unit
to $\varepsilon_n^\pm$,  

\noindent
then, the new formula so obtained
$\tilde \varepsilon_n^\pm$, will  formally be the same as the initial one but just  replacing  $n$ by $n+1$. This can be seen in the following scheme:
\begin{equation}
\varepsilon_n^\pm = n  \pm
\sqrt{\delta^2+n\lambda^2}  \ \implies \ 
\left\{\begin{array}{l} 
{\rm replace}\ \delta^2 \ {\rm by:}\ \delta^2+\lambda^2
\\[1.5ex]
{\rm add\ to}\  \varepsilon_n^\pm\ \ {\rm one\ unit:}\ \varepsilon_n^\pm+1
\end{array}\right\}
\ \implies \ \tilde \varepsilon_n^\pm  = n+1 \pm
\sqrt{\delta^2+(n+1)\lambda^2} = \varepsilon_{n+1}^\pm
\end{equation}
Then, if we start with the JC Hamiltonian $H_{\rm JC}(\delta)$ with parameter $\delta$,
  we can define a new almost isospectral JC Hamiltonian 
$\widetilde H_{\rm JC}^{(1)}(\delta)$  as follows:
\begin{equation}
H_{\rm JC}(\delta) \ \implies \ \widetilde H_{\rm JC}^{(1)}(\delta) :=H_{\rm JC}(\sqrt{\delta^2+\lambda^2}) +1
\end{equation}
One can easily check \cite{ates25}, that:

\begin{itemize}
\item The initial $H_{\rm JC}(\delta)$ and the new Hamiltonian
${\widetilde{H}}_{\rm JC}^{(1)}(\delta)$ are connected by means of an intertwining  operator $L$, given  by  
\begin{equation}
L= \left(\begin{array}{cc}
a^- & 0
\\[1.ex]
K & a^-
\end{array}\right)\,,\qquad K =\frac{\delta-\sqrt{\delta^2+\lambda^2}}{\lambda}
\end{equation}
where the intertwining relation reads
\begin{equation}
L\, H_{\rm JC}(\delta) = \widetilde H_{\rm JC}^{(1)}(\delta)\, L
\end{equation}
\item The spectrum of $\widetilde H_{\rm JC}^{(1)}(\delta)$ is the same as $H_{\rm JC}(\delta)$:  
\begin{equation}\label{iso}
\varepsilon^\pm_n=\tilde\varepsilon^\pm_{n-1}, \quad n=2,3,\dots
\qquad
\varepsilon^-_1 = \tilde\varepsilon^-_{0}
\end{equation}
except for the lowest two levels of the initial Hamiltonian: $\varepsilon^-_0$ and $\varepsilon^+_1$, which are missing in the new Hamiltonian (they are not included in (\ref{iso})).
\end{itemize}

\begin{figure}[h!]
\begin{center}
\includegraphics[width=0.45\textwidth]{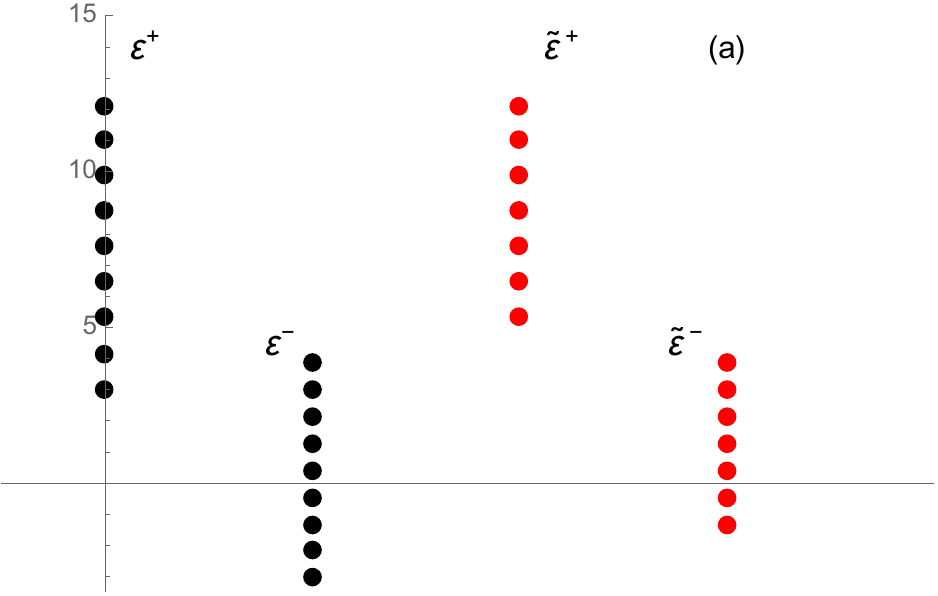}
\qquad\quad
\includegraphics[width=0.45\textwidth]{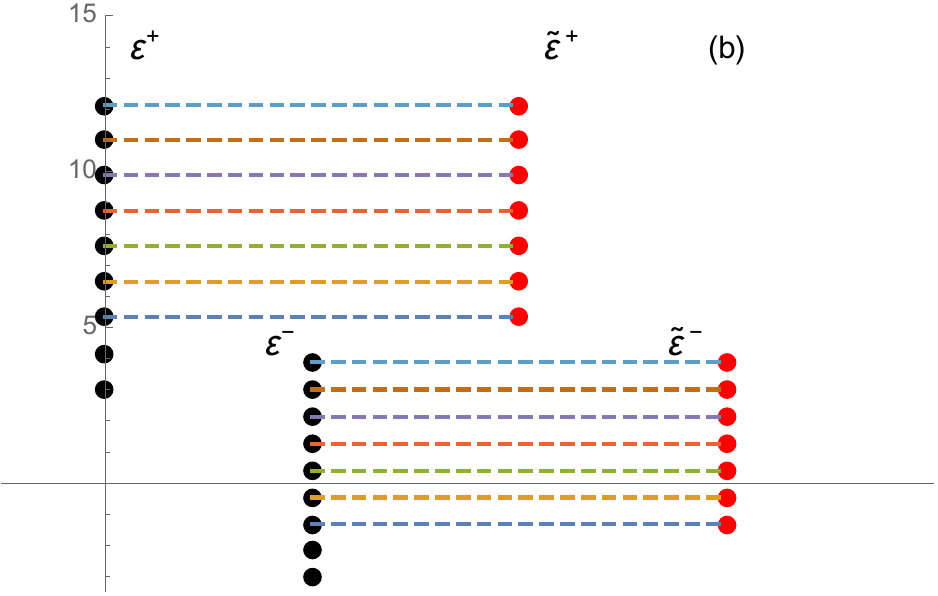}
\end{center}
\caption{\small (a) Plot of the first energy levels for both partner Hamiltonians: $H_{\rm JC}$ (in black) and $\tilde H_{\rm JC}^{(2)}$
(in red)  with
$\delta=3$, $\lambda=1$. (b) The common points of their  spectrum  are linked by dashing lines. Their spectra differ in 4  points, two of them from the `left spectrum' ($\varepsilon^+_1$ and $\varepsilon^+_2$) and two from the `right spectrum' 
($\varepsilon^-_0$ and $\varepsilon^-_1$) of $H_{\rm JC}$, each at the botton of its respective column. They do not have corresponding isospectral points in the partner Hamiltonian
$\tilde H_{\rm JC}^{(2)}$.
\label{fig2}}
\end{figure}

In the same way, by applying this process a number of consecutive times, from the initial JC Hamiltonian we can get a sequence of partner Hamiltonians,  which is called a hierarchy,
\begin{equation}\label{JCpn}
 \qquad \leftarrow\ {\widetilde{H}}_{\rm JC}^{(-k)}\  \leftarrow \dots \ {\widetilde{H}}_{\rm JC}^{(-1)}\ \leftarrow \ H_{\rm JC}(\delta) \ \to \ {\widetilde{H}}_{\rm JC}^{(1)}
\dots \ \to \ \ {\widetilde{H}}_{\rm JC}^{(k)}\ \to
\end{equation}
A generic $k$-SUSY partner in this hierarchy has the form
\begin{equation}\label{H3mk}
{\widetilde{H}}_{\rm JC}^{(k)}:=
\left(
\begin{array}{cc}  
a^- a^+ +k+\sqrt{\delta^2+k \lambda^2} &\lambda a^-
\\[1ex]
\lambda a^+ & a^+a^- +k - \sqrt{\delta^2+k \lambda^2}
\end{array}\right) 
\end{equation}
where $k \in \mathbb Z$. Its eigenvalues and (not normalized) eigenfunctions have the following expressions 
\begin{equation}\label{physk}
\begin{array}{lll}
\tilde \varepsilon_n^\pm = (n+k) \pm \sqrt{\delta^2 +(n+k) \lambda^2},\quad 
&\widetilde \Psi_n^\pm = \left(\begin{array}{c}
(\sqrt{\delta^2 + k \lambda^2}\pm \sqrt{\delta^2 + (n+k) \lambda^2})\,\psi_{n-1}
\\[1ex]
\sqrt{n}\,\lambda\, \psi_n
\end{array}\right),\quad &n=1,2,\dots
\\[4.5ex]
\tilde\varepsilon^-_0 = k-\sqrt{\delta^2 +k \lambda^2},\qquad
&\widetilde \Psi^-_0 = \left(\begin{array}{c}
0
\\[1ex]
 \psi_0
\end{array}\right),\quad &n=0
\end{array} 
\end{equation}
They are quite similar to the formulas (\ref{phys}) for the initial Hamiltonian.
The integer $k$ may be negative, but it is restricted by the condition $\delta^2 +k \lambda^2\geq 0$, {in order to have a Hermitian Hamiltonian operator}. In the case where  $\delta^2= \lambda^2  n_0$ (for a certain $n_0\in \mathbb N$), then the minimum value of $k$ will be $-n_0$. In that case, it happens that a JC Hamiltonian in resonance $\delta=0$ will start the hierarchy. 
An example of the spectrum of two SUSY partner Hamiltonians is given in Fig.~\ref{fig2}. We conclude that partner JC Hamiltonians are characterized by detuning parameters linked by previous formula: $\delta$ and  $\delta_k = \sqrt{\delta^2+ k \lambda^2}$. At the same time the coupling parameter $\lambda$ is kept unaltered.

{Another Hamiltonian that can be obtained by means of intertwining operators is called
the anti-JC Hamiltonian $H_{\rm aJC}$ \cite{ates25,negro24,lara05,solano03}. 
In this work, we will not consider anti-JC Hamiltonians, because the results are quite similar to those of standard JC Hamiltonians.}

\sect{Atomic inversion for SUSY partners of a JC Hamiltonian}

Let us consider the evolution of the initial state 
$\Psi_n=(\psi_{n},0)$. 
We want to find the evolution under any of the SUSY partner Hamiltonians $H_{\rm JC}^{(k)}$ mentioned in the above section. Firstly, we write the initial state as a linear combination of eigenfunctions of such Hamiltonian. Let us make use of the following notation: the normalized eigenfunctions obtained from (\ref{physk}) have the expressions
\begin{equation}
\widetilde \Psi_{n+1}^+= \left(\begin{array}{c}
c_n \psi_n
\\
s_n \psi_{n+1}
\end{array}\right)\,,\qquad
\widetilde \Psi_{n+1}^-= \left(\begin{array}{c}
s_n \psi_n
\\
-c_n \psi_{n+1}
\end{array}\right)
\end{equation}
where  $c_n, s_n $  satisfy $(c_{n})^2+(s_{n})^2=1$.  
Then, it is easy to check that
\begin{equation}\label{phys12}
\Psi_n=\left(\begin{array}{c}
\psi_{n}
\\[1ex]
0
\end{array}\right)=c_{n} \widetilde  \Psi_{n+1}^+ +s_{n}\widetilde  \Psi_{n+1}^- \quad n=1,2,\dots
\end{equation}

The time evolution operator $U(t)=e^{-i H_{\rm JC}^{(k)} t}$ acts on the solution given by Eq. (\ref{phys12}) according to their eigenvalues, as follows:
\begin{equation}\label{phys12t}
\Psi_n(t) =
c_{n} \widetilde  \Psi_{n+1}^+ e^{-i \tilde \varepsilon_{n+1}^+ t} 
+s_{n} \widetilde  \Psi_{n+1}^- e^{-i \tilde \varepsilon_{n+1}^- t}
\end{equation}

Then, the explicit time dependent spinor for a generic $k$-SUSY partner JC Hamiltonian $H_{\rm JC}^{(k)}$ is:
\begin{equation}\label{phitt}
\displaystyle \Psi_n(t,\delta)
=e^{-i (n+k+1) t}\left(\begin{array}{l}
 \left(\cos \left(t \sqrt{\delta ^2+\lambda ^2 (n+k+1)}\right)- \displaystyle \frac{i \sqrt{\delta^2+k\lambda^2}  \sin \left(t \sqrt{\delta ^2+\lambda ^2 (n+k+1)}\right)}{\sqrt{\delta ^2+\lambda ^2 (n+k+1)}}\right)\,\psi_n
\\[1ex]
- \displaystyle \frac{i \lambda  \sqrt{n+1}\,   \sin \left(t \sqrt{\displaystyle \delta ^2+\lambda ^2 (n+k+1)}\right)}{\sqrt{\delta ^2+\lambda ^2 (n+k+1)}}\,\psi_{n+1}
\end{array}\right)
\end{equation}
If we take $\delta=0$ and $k=0$, we can recuperate the resonant case \cite{gerry04}:
\begin{equation}\label{phittr}
\displaystyle {\Psi_n(t,\delta=0)}
=e^{-i (n+1) t}\left(\begin{array}{l}
 \cos \left(t \sqrt{\lambda ^2 (n+1)}\right)\,\psi_n
\\[1ex]
- \displaystyle {i  \sin \left(t \sqrt{\lambda ^2 (n+1)}\right)}\,\psi_{n+1}
\end{array}\right)
\end{equation}
As a first stage, let us consider the atomic inversion evolution of this state which is given by
\begin{equation}\label{wn}
W_n(t)={\langle\Psi_n(t,\delta)|\sigma_3|\Psi_n(t,\delta)\rangle}
\end{equation}
The result is
\begin{equation}\label{wnf}
W_n(t,k,\delta)=\frac{\delta ^2+k \lambda^2+\lambda ^2 (n+1) \cos \left(2 t \sqrt{\delta ^2+\lambda ^2 (n+k+1)}\right)}{\delta ^2+\lambda ^2 (n+k+1)}
\end{equation}
If we choose $k=0$ and $\delta=0$, we  recuperate the atomic inversion function for resonance case $W_n(t,k=0,\delta=0)= \cos \left(2 t \sqrt{\lambda ^2 (n+1)}\right)=\cos \Omega (n)t$ \cite{gerry04},
where the generalized Rabi frequency for the $k$ partner is
\begin{equation}
\Omega(n,\delta)=\varepsilon_{n+1}^+-\varepsilon_{n+1}^-=2\sqrt{\delta^2+(n+k+1)\lambda^2}
\end{equation}

Next, we will consider the atomic inversion for a JC Hamiltonian partner when we consider as initial state a coherent state $
\psi_{\alpha}:=|{\alpha}\rangle
$,
and the atom is in an excited state.
In our case a coherent state is defined as a linear combination of number states such that it satisfies the relation $a^-|{\alpha}\rangle=\alpha|{\alpha}\rangle$ with a complex $\alpha$.
This type of coherent Glauber states has the expression \cite{gerry04} where $|m\rangle$ represents number state:
\begin{equation}\label{coh}
|{\alpha}\rangle=\sum_{m=0}^\infty C_m |m\rangle, \qquad  C_m= e^{-|\alpha|^2/2}\frac{\alpha^m}{\sqrt{m!}}
\end{equation}
The mean value of the number operator $N=a^+ a^-$  in a coherent state is $\overline{n}= |\alpha|^2$.
Thus, we will take the initial state
given in terms of this coherent state by
\begin{equation}
{\Psi_\alpha}= \left(\begin{array}{l}
1
\\[1.ex]
0
\end{array}\right)\otimes |\alpha\rangle = 
\left(\begin{array}{l}
|\alpha\rangle
\\[1.ex]
0
\end{array}\right) = \sum_{m=0}^\infty C_m \Psi_m
\end{equation}
where $\Psi_m$ were defined in (\ref{phys12}) and (\ref{phys12t}). Therefore,
the time evolution  is as follows:
\begin{equation}
{\Psi_\alpha(t)}= 
\sum_{m=0}^\infty C_m \Psi_m(t)
\end{equation}
with $\Psi_m(t)$ given by (\ref{phitt}). The evolution of the atomic inversion (\ref{wnf}) of the  coherent-type initial state (\ref{coh}) will be
\begin{equation}\label{wa}
W_\alpha(t,k,\delta)= \sum_{m=0}^\infty |C_m|^2 W_m(t,k,\delta)
\end{equation}

\subsection{Classical and revival times for atomic inversion of  JC-partners}

\begin{figure}[h!]
\begin{center}
\includegraphics[width=0.4\textwidth]{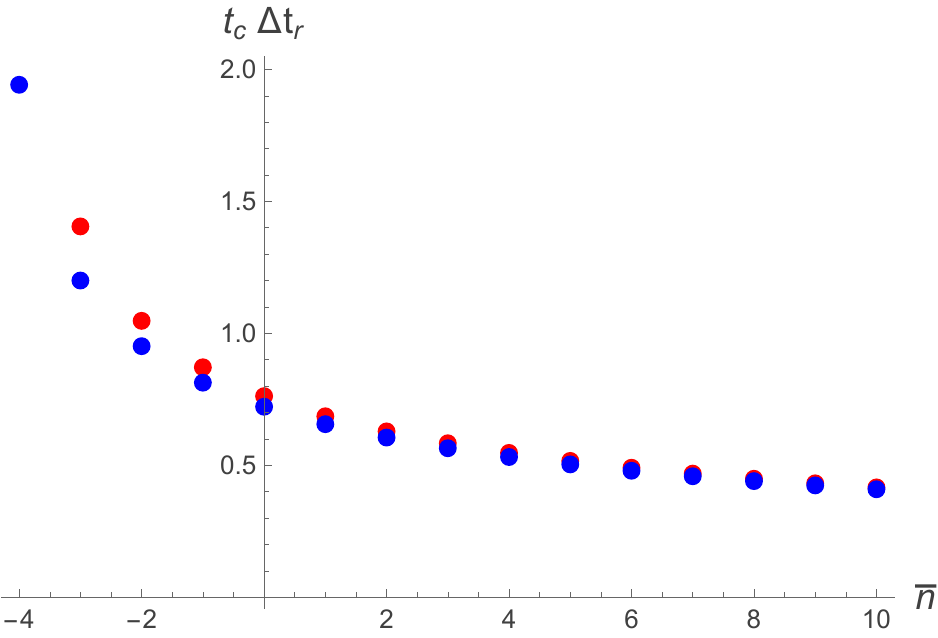}
\end{center}
\caption{\small Red one is classical times  and
blue one is  differences of revival times for $\tilde H_{\rm JC}^{(k)}$ for $\delta=3,\lambda=1, \bar n=4$, $k=-13,\dots,10$.
\label{fig3}}
\end{figure}

Before plotting the evolution of the time inversion, we will compute the relevant times from the Rabi frequency (\ref{wnf}) in the case of the evolution of one Fock state $|n\rangle$ and formula (\ref{wa}) for the evolution of a coherent state. 
From (\ref{wnf}) we define the classical time \cite{robinett04} for the initial function $\Psi_n$ as:
\begin{equation}\label{t0}
t_{\rm c} =\frac{\pi}{\sqrt{\delta^2+(\bar n+k+1)\lambda^2}}
\end{equation}
In the case of the evolution (\ref{wa}) of {$\Psi_\alpha(t)$}, we should define the classical time   
replacing $n$ by $\overline{n} = |\alpha|^2$.
While the revival time \cite{robinett04}  for  { $\Psi_\alpha(t)$ }is defined by
\begin{equation}\label{t0}
{t_{\rm r}} = \pi \frac1{\frac{d}{{d {n}}}
\sqrt{\delta^2+({n} +k+1)\lambda^2}}\big|_{n = \bar n}
=
\frac{2\pi \sqrt{\delta^2+(\overline{n} +k+1)\lambda^2}}{\lambda^2}
\end{equation}
where $\overline{n}= |\alpha|^2$. The revival and classical times {$ t_{\rm r}$  and $t_{\rm c}$ depend on the variables $\delta, k, \lambda$ and  $\overline{n}$, but in order to simplify the notation, we will omit then.  When necessary, we will put explicitly only the relevant variable.
We plot these classical and revival times for some examples of SUSY partners of the JC Hamiltonian in  Fig.~\ref{fig3}.
It can be checked that that the difference of revival times given of consecutive partner JC Hamiltonians are equal to the classical time 
$t_{\rm c}(k)$
\begin{equation}\label{trtc}
\Delta t_{\rm r}= t_{\rm r}(k)-t_{\rm r}(k-1)\approx \partial t_{\rm r}(k)/\partial k = t_{\rm c}(k)\qquad {\rm at}\qquad \bar n
\end{equation}
The fact that the difference of  time revivals for consecutive partner Hamiltonians coincide with the classical time, means that the time interval between two maxima of consecutive SUSY partners at $t_{\rm r}$ will have a period of classical oscillations, which implies that these maxima will be in phase (see Fig. 5). However, when we look at the maxima of a JC Hamiltonian and that of a non SUSY partner, they will have a separation different to that of the classical $t_{\rm c}$. Therefore, these two maxima will appear not in phase (see Fig. 6). In conclusion, a track of the SUSY partners is that the oscillations near each maxima of each revival time will almost coincide with the maxima (or minimum) values. The plots of Figs. 4 and 5 for three consecutive SUSY partners illustrate this behavior, while Fig. 6 shows clearly how different is for a non SUSY partner with respect to these three SUSY partners.}

\begin{figure}[h!]
\begin{center}
\includegraphics[width=0.5\textwidth]{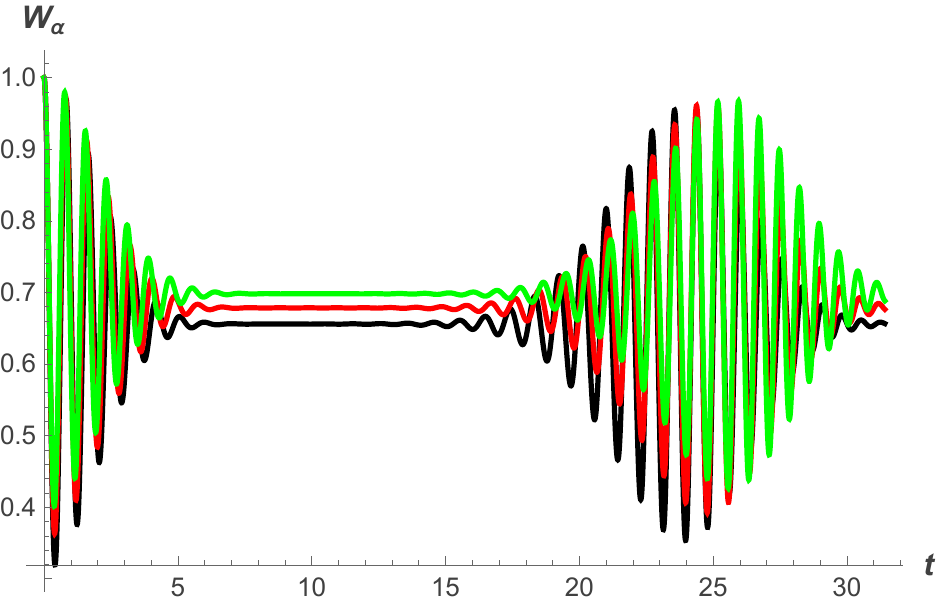}
\end{center}
\caption{\small Evolution of atomic inversion for a JC Hamiltonian
with $\delta=3, \lambda=1, \alpha=2$ (black) and two next SUSY partners (for $k=1$ in red and for $k=2$ in green).
\label{fig4}}
\end{figure}

\begin{figure}[h!]
\begin{center}
\includegraphics[width=0.4\textwidth]{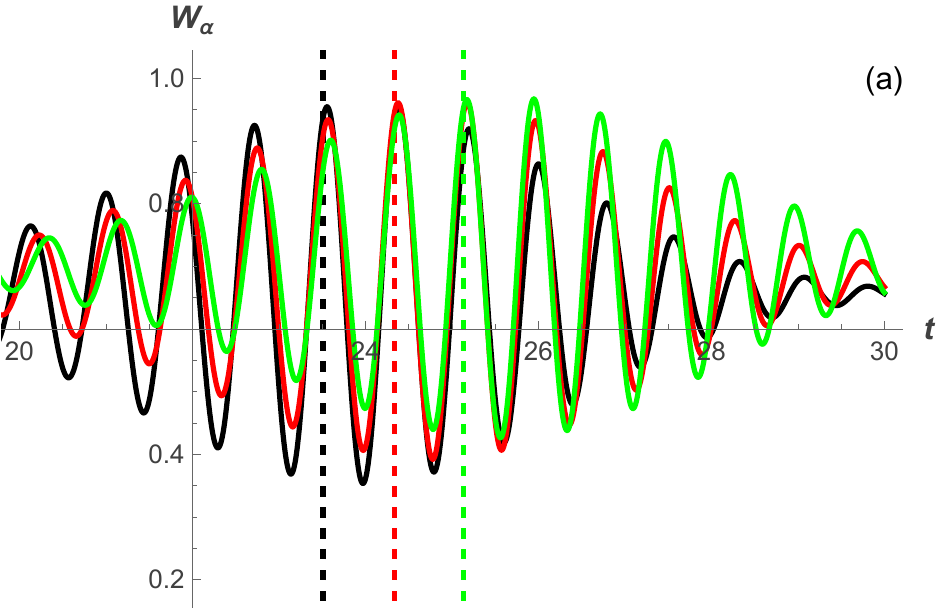}
\qquad\quad
\includegraphics[width=0.35\textwidth]{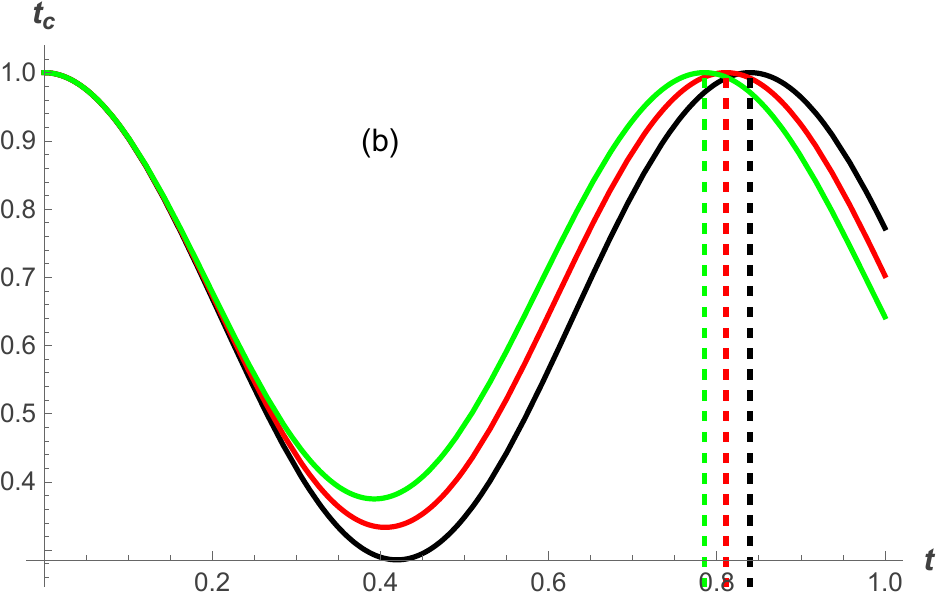}
\end{center}
\caption{\small Details of Fig.~\ref{fig4}. (a) shows that the revival  times given by dashed vertical lines coincide with the maxima of the oscillations.   (b) shows the classical times for the three cases.
\label{fig5}}
\end{figure}

In Fig.~\ref{fig4}, we plot the atomic inversion for one JC Hamiltonian
together with two consecutive partner Hamiltonians.
In   Fig.~\ref{fig5}, we can see some specific details  of Fig.~\ref{fig4} concerning
classical and revival times: it is shown that the difference of revival times of consecutive partners is approximately equal to the classical time. This is easily appreciated because they are `in phase'.

\begin{figure}[h!]
\begin{center}
\includegraphics[width=0.5\textwidth]{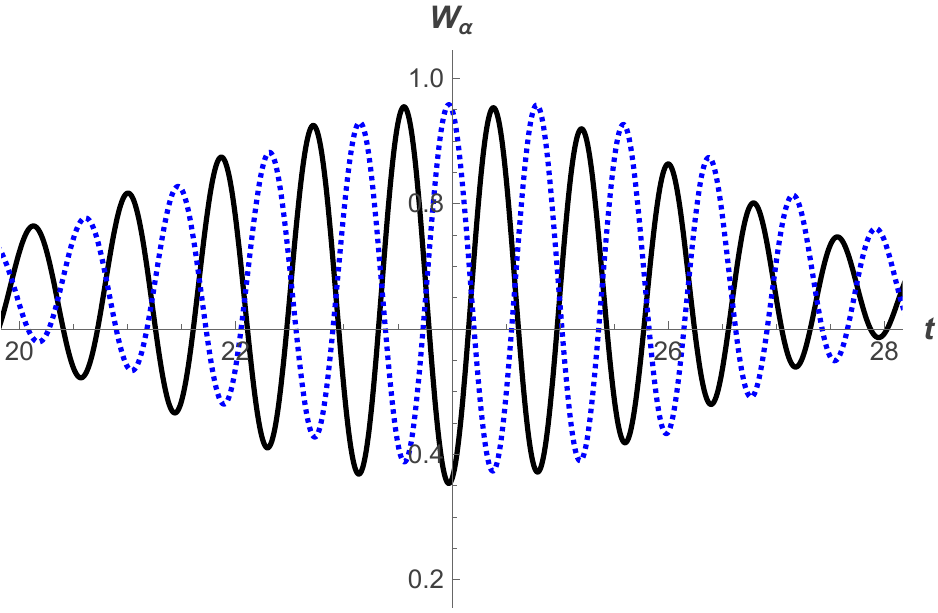}
\end{center}
\caption{\small Atomic inversion for a JC Hamiltonian: ($\delta=3, \lambda=1, \alpha=2$) (black) and a non SUSY partner JC Hamiltonian:  ($\delta=\sqrt{9.5}, \lambda=1, \alpha=2$) (in blue). 
\label{fig6}}
\end{figure}

{We can appreciate the difference of SUSY partners evolution of Fig.~\ref{fig5} with the evolution of non SUSY partners of Fig.~\ref{fig6} due to a change of the detuning $\delta$ parameter.  
We conclude that expected values of atomic inversion operator characterize quite well the partnership of JC Hamiltonians.} 

\sect{Expected values of quadrature operators}

We want also to examine the expectation values of pure field operators such as the creation and annihilation operators $a^+$ and $a^-$ of the quantum radiation field with frequency $\omega$. The first important difference is that since the evolution of $a^\pm$ takes complex values,
therefore this suggests a plot of the results by the evolution in time of both,  real and imaginary parts of their expected values, { $\langle \Psi_\alpha(t)| a^\pm \Psi_\alpha(t)\rangle$}.
We also compare these parametric values with the well known expected values of $a^\pm$ in a coherent state $|\alpha(t)\rangle$ of the free radiation field $H_f = a^+ a^-$ which are given by
{$\langle \alpha(t)| a^- \alpha(t)\rangle= \alpha e^{-i \omega t} $
and $\langle \alpha(t)| a^+ \alpha(t)\rangle=\langle a^-\alpha(t)|\alpha(t)\rangle= \alpha^* e^{i \omega t} $. }

In this case, the expected value contains terms with different frequencies. For instance, there appear two types of periods in { $\langle \Psi_\alpha(t)| a^\pm \Psi_\alpha(t)\rangle$} given by:
\begin{equation}
\left\{\begin{array}{l}
{\omega_1(\overline n)= 1+\sqrt{\delta^2+(k+\overline n) \lambda^2} + \sqrt{\delta^2+(k+\overline n+1) \lambda^2}}
\\[1.5ex]
{\omega_2(\overline n)= 1+ \sqrt{\delta^2+ (k+\overline n) \lambda^2} - \sqrt{\delta^2+(k+\overline n+1) \lambda^2}}
\end{array}
\right.
\end{equation}
From these frequencies we find two main periods:
\begin{equation}
\left\{
\begin{array}{l}
{T_{c} = \frac{2\pi}{\omega_2}}
\\[1.5ex]
{T_{r}  = {2\pi}/{\frac{d\, \omega_1}{d \overline n}}}
\end{array}
\right.
\end{equation}
But, in this case, there is no clear relation between the classical and revival time as it was the case for the atomic inversion.
\begin{figure}[h!]
\begin{center}
\includegraphics[width=0.4\textwidth]{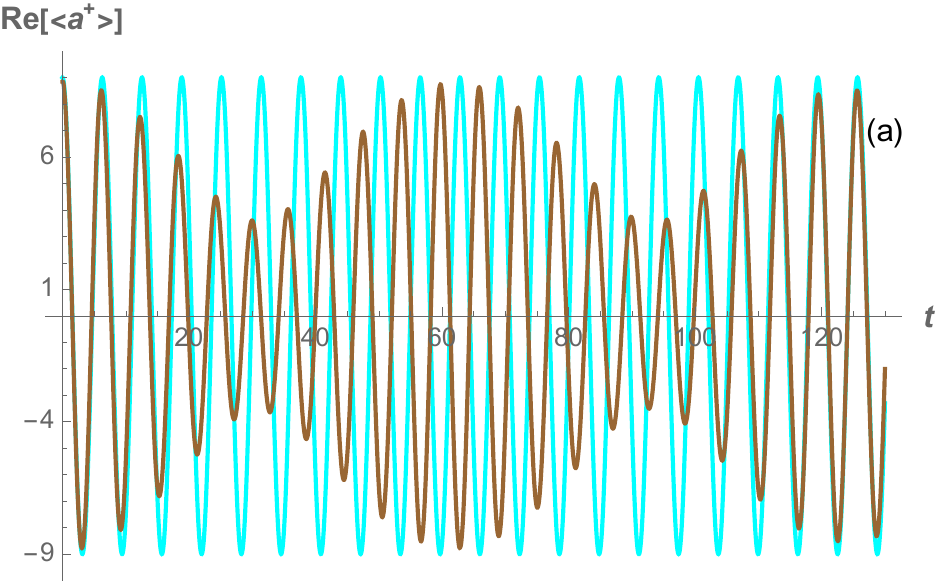}
\qquad
\includegraphics[width=0.4\textwidth]{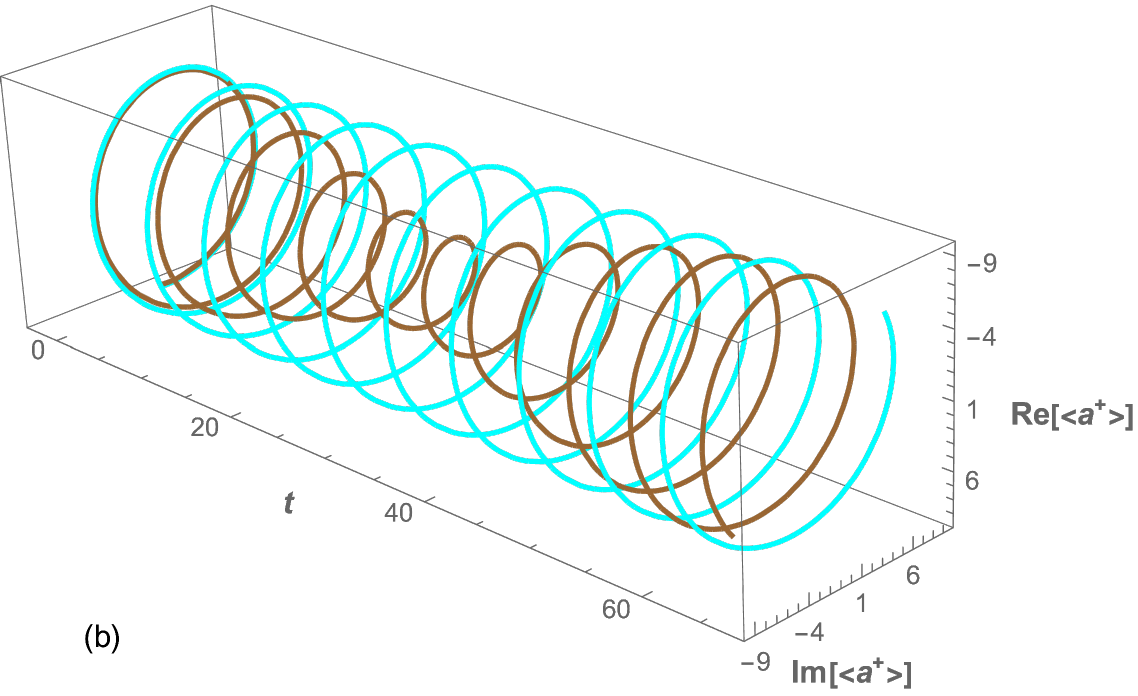}
\end{center}
\caption{
\small (a) Expected values  of the real part  
 {$\langle \Psi_\alpha(t)| a^+\Psi_\alpha(t)\rangle$} for the operator $a^+$  together with the free radiation evolution
{$\langle \alpha(t)| a^+\alpha(t)\rangle$} for $(\omega=1, \delta=4,\lambda=1, \alpha=9)$. (b) We plot the real and imaginary parts of these expected values.
\label{fig7}}
\end{figure}
\begin{figure}[h!]
\begin{center}
\includegraphics[width=0.4\textwidth]{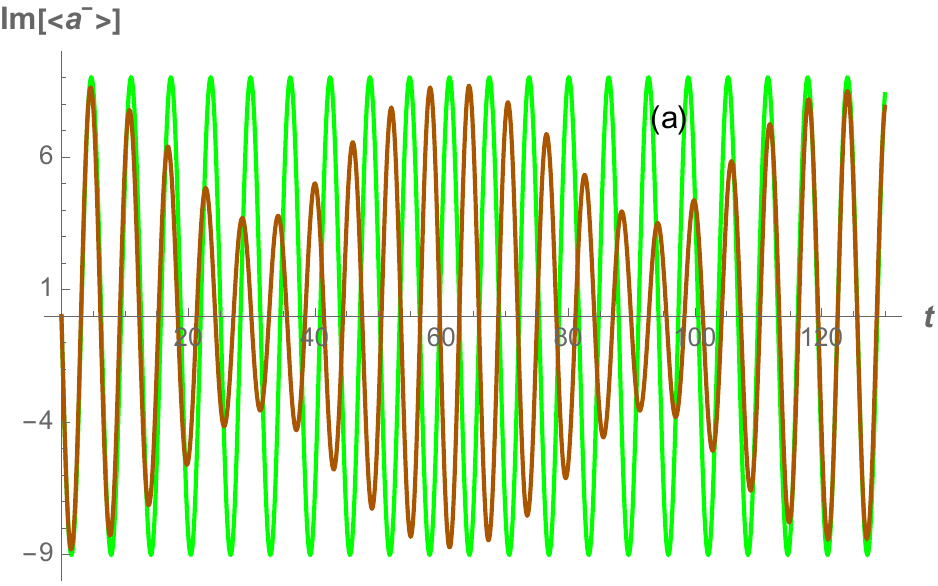}
\qquad
\includegraphics[width=0.4\textwidth]{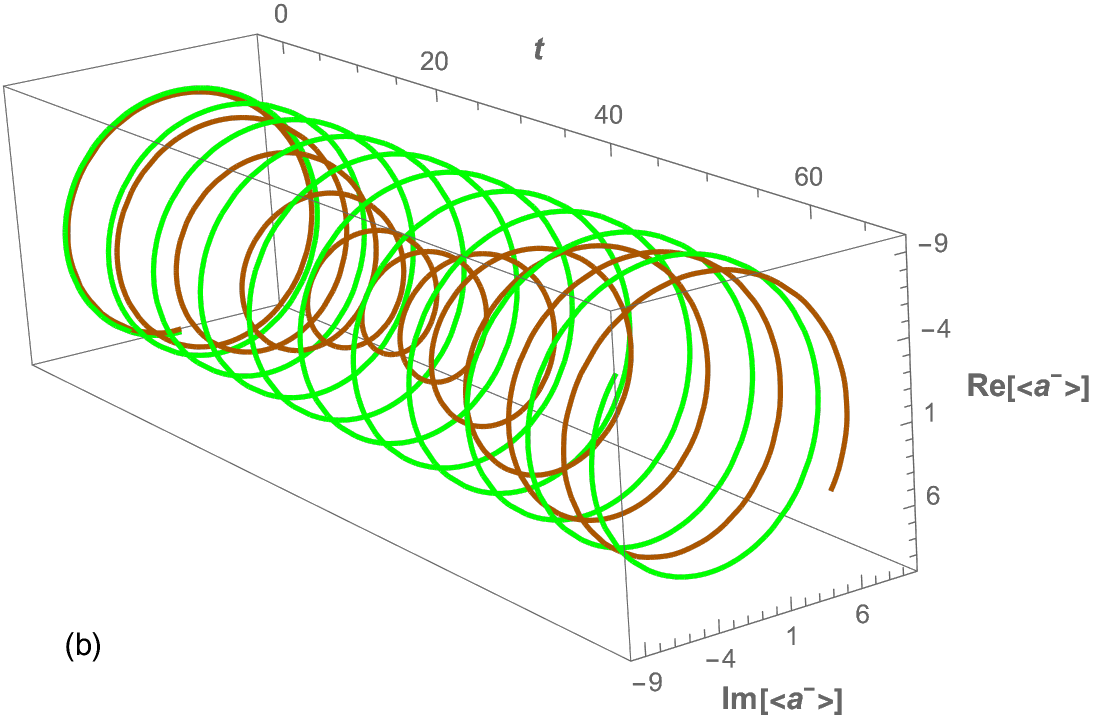}
\end{center}
\caption{
\small (a) Expected values  of the imaginary part  
{$\langle \Psi_\alpha(t)| a^-\Psi_\alpha(t)\rangle$} of the operator $a^-$  together with the free radiation case
{$\langle \alpha(t)| a^-\alpha(t)\rangle$} $(\omega =1, \delta=4,\lambda=1, \alpha=9)$. (b) We plot both real and imaginary parts of these expected values.
\label{fig8}}
\end{figure}

\begin{figure}[h!]
\begin{center}
\includegraphics[width=0.4\textwidth]{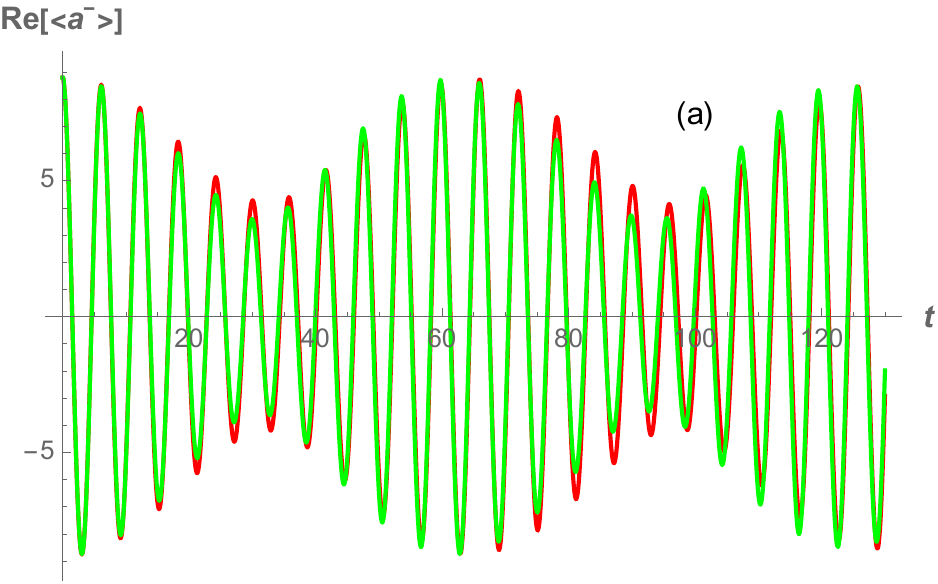}
\qquad
\includegraphics[width=0.4\textwidth]{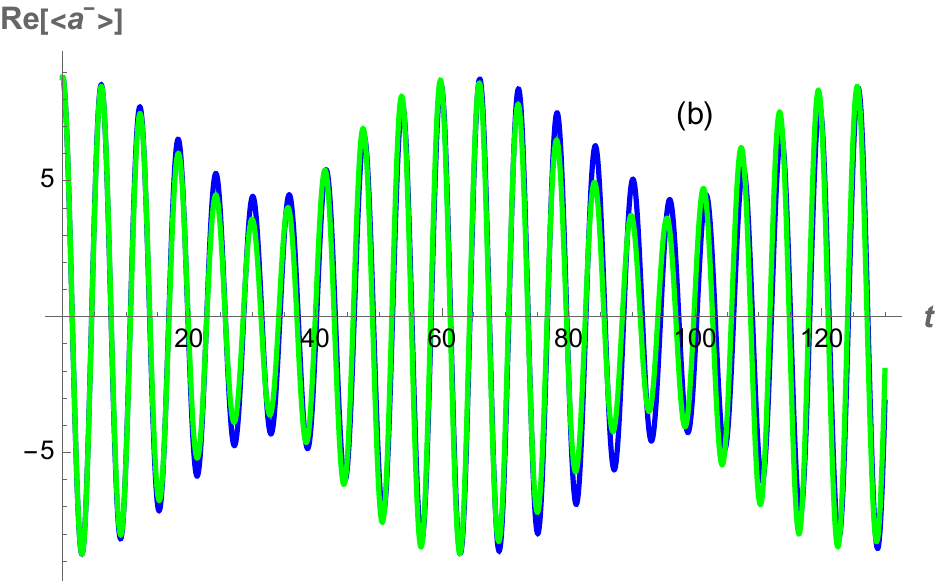}
\end{center}
\caption{
\small (a) Expected values  of the real part  
{$\langle \Psi_\alpha(t)| a^-\Psi_\alpha(t)\rangle$} with parameters 
$(\omega =1, \delta=4, \lambda=1, \alpha=9)$ for the JC Hamiltonian (green) and its 8th SUSY partner (red)  (b) The same comparison between the evolution under the initial Hamiltonian (green) and a non-SUSY partner with $(\delta=5.2,\lambda=1, \alpha=9)$ ( blue).
\label{fig9}}
\end{figure}

\begin{figure}[h!]
\begin{center}
\includegraphics[width=0.4\textwidth]{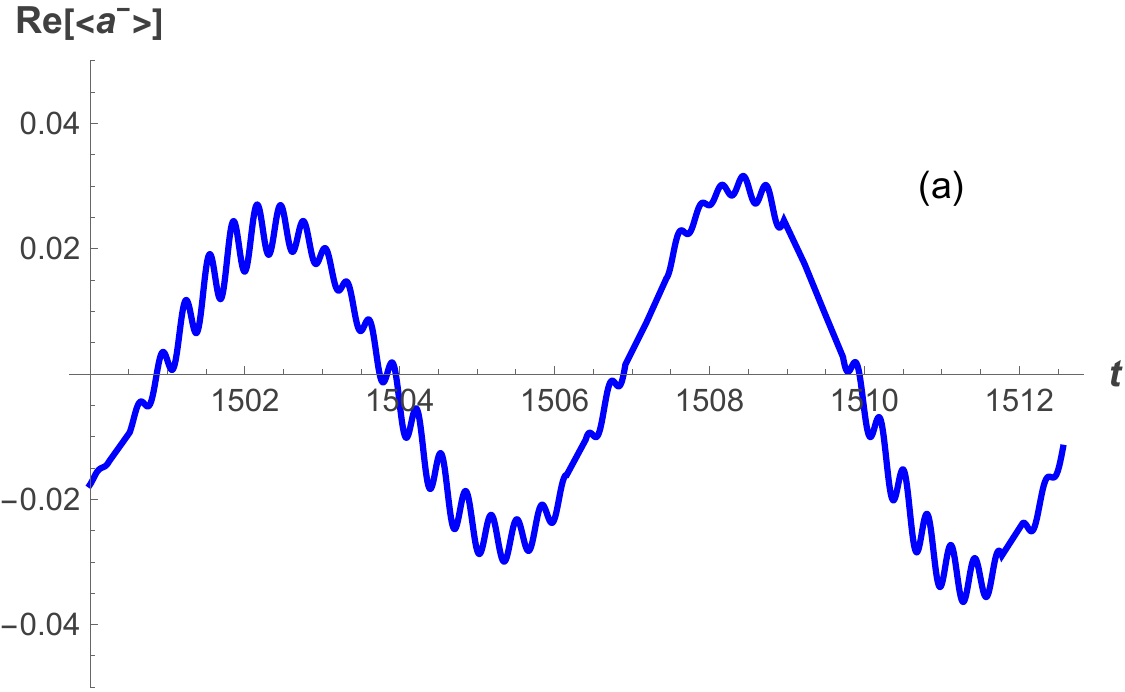}
\qquad
\includegraphics[width=0.4\textwidth]{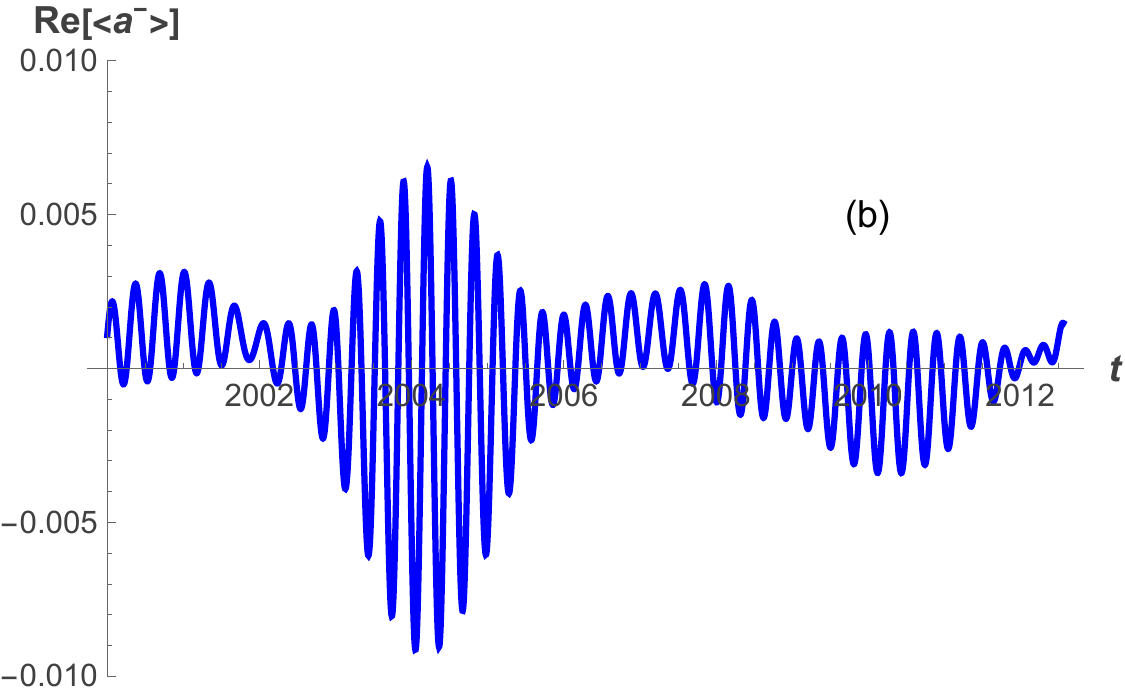}
\\
\includegraphics[width=0.4\textwidth]{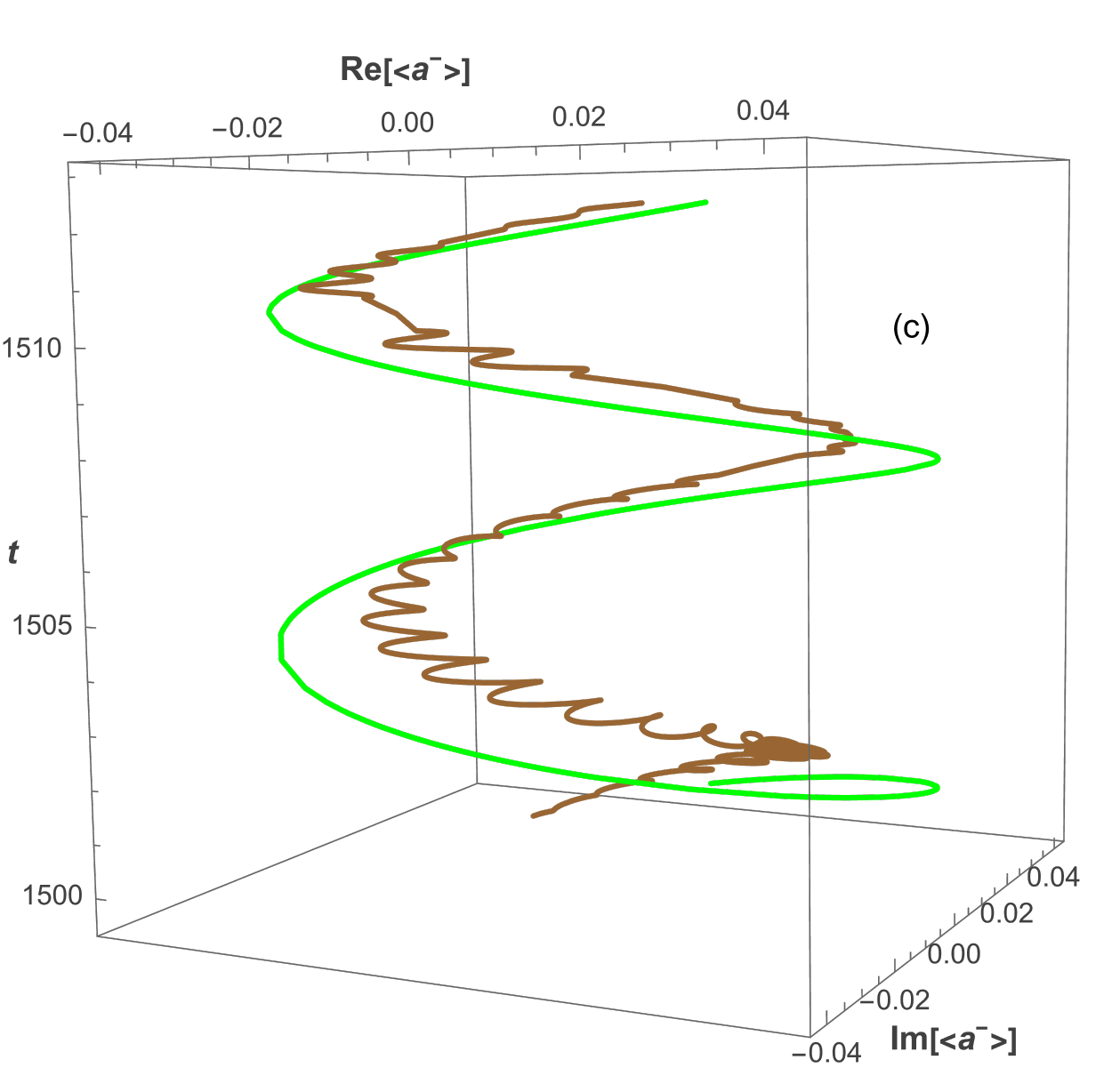}
\qquad
\includegraphics[width=0.4\textwidth]{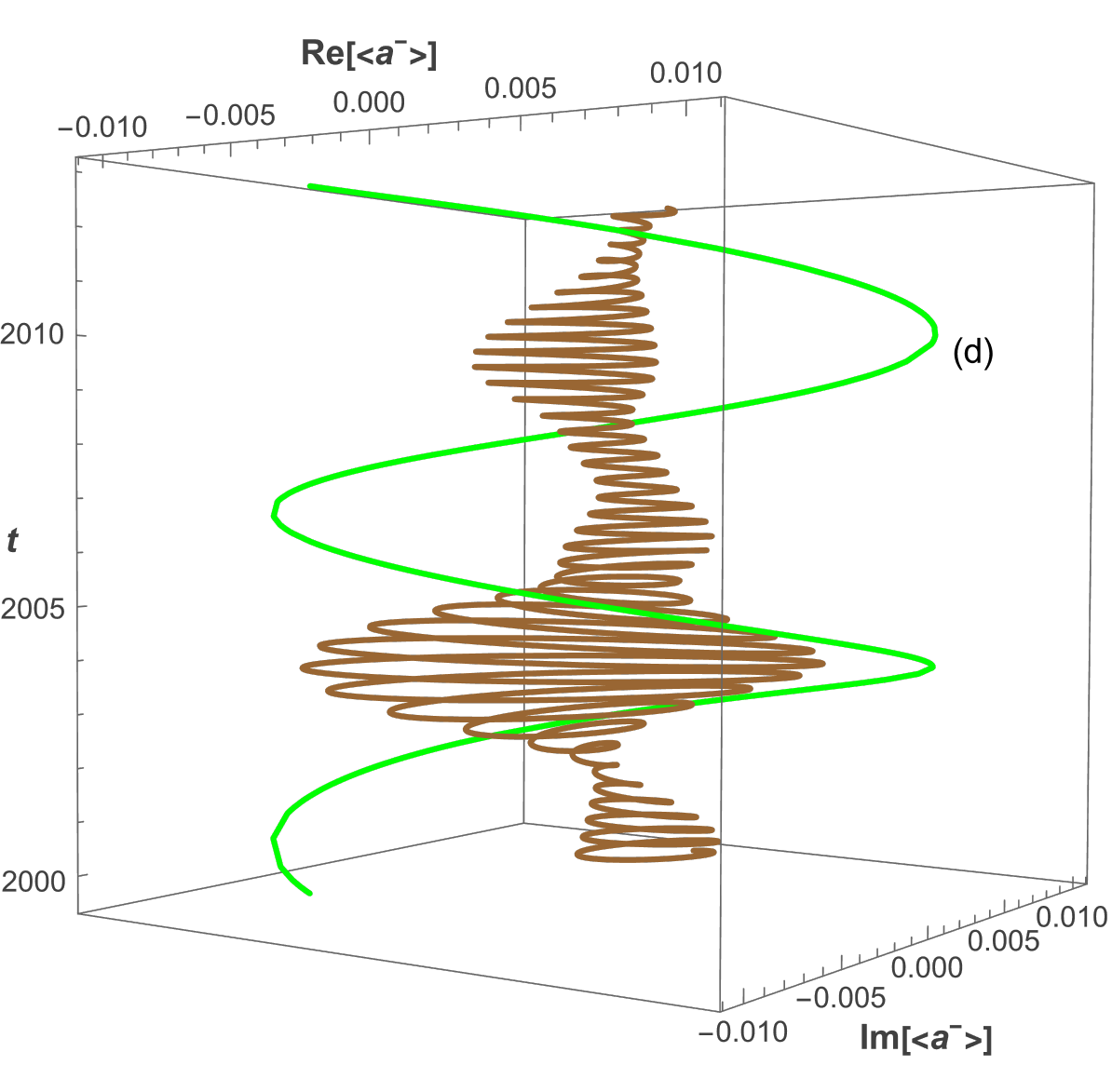}
\end{center}
\caption{{\small (a) 2D plot of the expected value
$\text{Re}(\langle \Psi_\alpha(t)| a^-\Psi_\alpha(t)\rangle)$ in the time interval $(1500,1500+4\pi)$;
(b) 2D plot of the expected value
$\text{Re}(\langle \Psi_\alpha(t)| a^-\Psi_\alpha(t)\rangle)$ in the time interval $(2000,2000+4\pi)$; 
(c) 3D plot of the expected value
$(\text{Re}(\langle \Psi_\alpha(t)| a^-\Psi_\alpha(t)\rangle),\text{Im}(\langle \Psi_\alpha(t)| a^-\Psi_\alpha(t)\rangle))$ in the time interval $(1500,1500+4\pi)$; 
(d) 3D plot of the expected value
$(\text{Re}(\langle \Psi_\alpha(t)| a^-\Psi_\alpha(t)\rangle),\text{Im}(\langle \Psi_\alpha(t)| a^-\Psi_\alpha(t)\rangle))$ in the time interval $(2000,2000+4\pi)$; 
All the plots have parameter values
$(\omega=1, \delta=4,\lambda=1, \alpha=9, \theta=0)$. The green curve is that of a classical frequency expected for a coherent state,  taken as reference.}
\label{fig10}}
\end{figure}


\begin{figure}[h!]
\begin{center}
\includegraphics[width=0.4\textwidth]{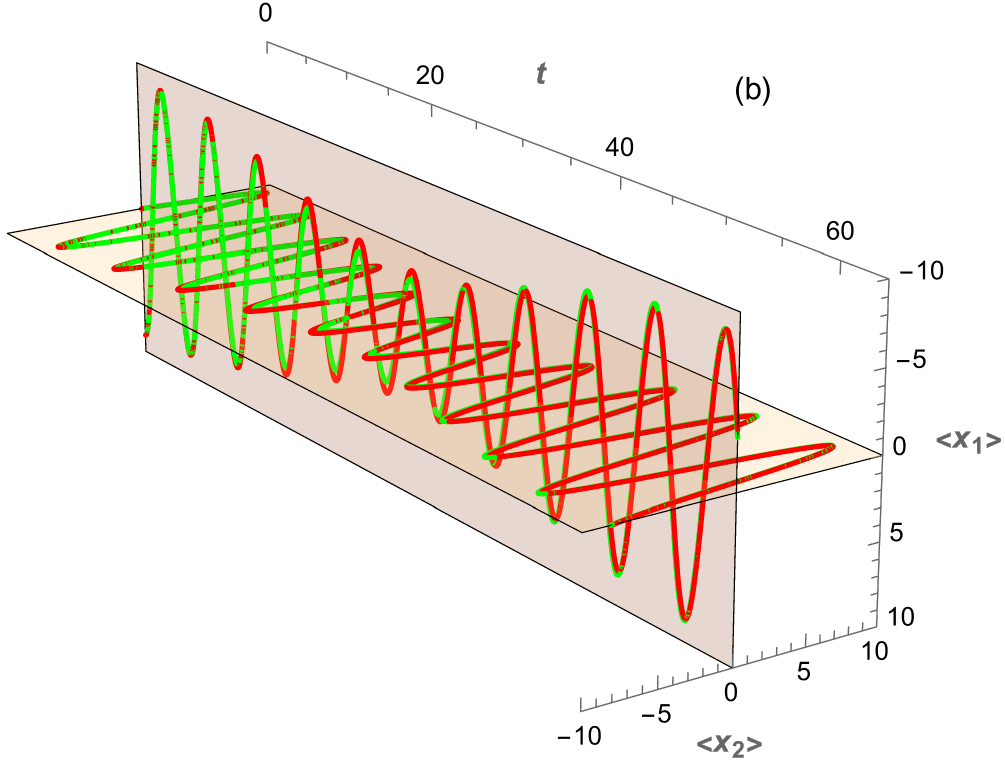}
\end{center}
\caption{
\small Expected values, 
inside two perpendicular planes, of the quadrature operators $x_1= (a^+ + a^-)/2$ and $x_2= (a^- - a^+)/(2i)$ for the values $(\omega =1, \delta=4, \lambda=1, \alpha=9)$. The evolution under  the JC Hamiltonian $ H_{\rm JC}$ is in green, while red of the evolution under a partner $\widetilde H_{\rm JC}^{(8)}$. 
\label{fig11}}
\end{figure}

{Although the creation and  annihilation operators, $a^+$ and $a^-$, are not observables, these operators take part of some physical observables. For example, the quadratures ($x_1=(a^+ + a^-)/2, x_2 = (a^- - a^+)/2i $), or the number operator $N$. Therefore, it is worth to see the behavior of their expectation values to understand better what happens with observables.} Figs.~\ref{fig7}-\ref{fig8} represent the evolution of expected values of
 $a^+$ and $a^-$ for the state {$\Psi_\alpha(t)$}  under a JC Hamiltonian (in Figs.~\ref{fig7}a and \ref{fig8}a the real part of the expected value, while Figs.~\ref{fig7}b and \ref{fig8}b both real and imaginary parts). It is also included the evolution of the expected value in a coherent state (in cyan color), so that we can appreciate the how is the evolution under a JC Hamiltonian. In Figs.~\ref{fig7}-\ref{fig8}  the oposite sign in the frequency of {$\langle \Psi_\alpha(t)| a^{\pm}\Psi_\alpha(t)\rangle$} is made explicit. We can say that the evolution of {$\langle \Psi_\alpha(t)| a^{\pm}\Psi_\alpha(t)\rangle$} under the JC Hamiltonian is surprisingly quite similar to a coherent state, at least after a few periods of revival time.  { 
 For greater times, we have noticed that the expected values of $a^-$ changes, reproducing not pure coherent curves but apparently including other frequencies that differ from those of the coherent states (see Fig.~\ref{fig10}a and Fig.~\ref{fig10}c) with associated revivals (see Fig.~\ref{fig10}b and Fig.~\ref{fig10}d). In the long time, we have observed that the intensity of the revivals of the expected value of $a^-$ decreased as was described by \cite{eberly80}. Therefore, the expected values of the field operator present new interesting properties at sufficiently long times. The general behaviour at large time follows the same features as other operators described for example in references \cite{eberly80, moya93, moya94}.}

{Next, we show the evolution of these field operators under two JC Hamiltonians  which are SUSY partners and two which are not SUSY partners in Fig.~\ref{fig9}. We observe that there is no significant difference between being SUSY partner or not. In conclusion, expected values of field operators are not appropriate to differentiate the SUSY partnership. This is in contrast to expected values of atomic inversion operator. This may be due to the evolution of expected values of field operators involve many frequencies, which eliminate sensible measures of the different behaviour.}


In a second step, we have considered also the evolution of the expected values of the quadrature operators $x_1= (a^+ + a^-)/2$ and 
 $x_2= (a^- - a^+)/(2i)$. These operators are Hermitian, and therefore, their expected values are real. One of them describes the evolution of a `position operator', while the other is that of  `momentum'. 
A representation of both quadrature operators is given in Fig.~\ref{fig11}. 
We have also included the evolution, when we choose a partner JC Hamiltonian and check that the behaviour follows the same track as the original $H_{\rm JC}$  Hamiltonian but again there are no significant differences between SUSY and non SUSY partners.

\sect{Conclusions}

Our aim in this paper was to show the specific behaviour on the evolution of  expected values such as  atomic inversion, operators $a^\pm$ and quadratures under SUSY partner Hamiltonians $\widetilde H^{(k)}_{\rm JC}$,
$k\in \mathbb Z$.  SUSY partner Hamiltonians have the same spectrum except for a finite number of the lowest levels (they are almost isospectral), so this property could have some kind of influence on the values of expectations values.

The first type of expected values was the atomic inversion $\langle \sigma_3\rangle$.
We computed some features of the evolution such as the classical times and revival times. It was clear that the evolution of atomic inversion is very sensitive to SUSY partners. Non SUSY partner Hamiltonians present qualitative differences. This is explicit in the two representative plots for non SUSY partners of Fig.~\ref{fig6} compared to those of Figs.~\ref{fig4}-\ref{fig5} corresponding to SUSY partners.

The second type of expected values concern to the field operators $a^\pm$. We have computed the relevant classical and revival times and we have plotted the evolution under JC Hamiltonian and the usual coherent state evolution. {We have preferred to plot  the expected values of these operators in parametric type form, because the values are complex. So, it is natural to represent these values in a three dimensional space, with one axis for time and the perpendicular plane for the complex values as shown in Figs.~\ref{fig7}b-\ref{fig8}b. One can apreciate what it is going on: the modulus has colapses/revivals, while the frequency of the evolution also suffer a type of colapse/revival evolution.} The coherent behaviour of $\langle a^\pm\rangle$ is conserved along  a few revival periods, where there are slight changes in amplitude and frequency. However,  at very longer times the interference effects dominate and new interesting properties occur (see this evolution in Figs.~\ref{fig10}). These parametric 2D and 3D plots can be complemented with polar type plots as shown for instance, in Ref.~\cite{lara21}, where it is quite different the information they supply.  

In this case, the evolution of the expected values under SUSY partners is similar to non SUSY partners as it is shown in Figs.~\ref{fig9}a and \ref{fig9}b. So in this case, we conclude that this type of measures do not differentiate SUSY partners from non SUSY partnership. This is due to very complex time evolution of this kind of expected values where there appear two fundamental frequencies. We also mention that since expected values of quadratures are real, then they allow to be plotted in two perpendicular planes as shown in Fig.~\ref{fig11}. In this case both SUSY and non SUSY partners  evolution is quite coherent up to a certain time, but again they are quite similar and there is no a special behaviour which could characterize SUSY partnership.

{In principle it is possible to consider the building of SUSY transformations for other similar systems. For example the Tavis-Cummings model does not have anti-rotating terms, so this type of methods can be applied. The case of Dicke would have similar difficulties as the quantum Rabi model what makes difficult to find intertwining operators.}

\section*{Data availability statement}
No new data were created or to be shared in this study.

\section*{Acknowledgments}

The authors would like to acknowledge the support of the project PID2023-148409NB-I00, funded by MICIU/AEI and the contribution of the European Cooperation in Science and Technology COST Action CA23130.
\c{S}.~Kuru thanks Ankara University and the warm hospitality of the Department of Theoretical Physics of the University of Valladolid, where part of this work has been carried out, and to the support of its GIR of Mathematical Physics.  \.{I}. B.~Ate\c s thanks to TUBITAK BIDEB 2211-A program.

\end{document}